\pdfoutput=1
\documentclass[aps,prl,reprint,showpacs,showkeys,preprintnumbers,superscriptaddress]{revtex4-1}

\usepackage{color}
\usepackage{cancel}
\usepackage[colorlinks=true
,urlcolor=magenta
,anchorcolor=green
,citecolor=red
,filecolor=green
,linkcolor=blue
,menucolor=green
,linktocpage=true
,pdfauthor={Christian Ecker, Daniel Grumiller, Philipp Stanzer, Wilke van der Schee}
,pdfa=true
]{hyperref}
\usepackage{graphicx,epsfig,color}
\usepackage[english]{babel}
\usepackage{amssymb,amsfonts,amsmath}
\usepackage{verbatim}

\newcommand{\eq}[2]{\begin{equation} #1 \label{#2} \end{equation}}
\DeclareMathOperator{\extdm}{d}
\newcommand{\extd}{\extdm \!}

\newcommand{\be}{\begin{equation}} \newcommand{\ee}{\end{equation}}
\newcommand{\bea}{\begin{eqnarray}} \newcommand{\eea}{\end{eqnarray}}
\newcommand{\beann}{\begin{eqnarray*}}  \newcommand{\eeann}{\end{eqnarray*}}
\newcommand{\bfig}{\begin{figure}} \newcommand{\efig}{\end{figure}}
\newcommand{\ba}{\begin{array}} \newcommand{\ea}{\end{array}}
\newcommand{\bcen}{\begin{center}} \newcommand{\ecen}{\end{center}}
\newcommand{\btab}{\begin{tabular}} \newcommand{\etab}{\end{tabular}}

\newtheorem{Proposition}{Proposition}[section]

\newtheorem{Theorem}{Theorem}[section]
\newtheorem{Lemma}{Lemma}[section]
\newtheorem{Corollary}{Corollary}[section]

\newcommand{\bp}{\begin{Proposition}}   \newcommand{\ep}{\end{Proposition}}
\newcommand{\bt}{\begin{Theorem}}   \newcommand{\et}{\end{Theorem}}
\newcommand{\bl}{\begin{Lemma}}     \newcommand{\el}{\end{Lemma}}
\newcommand{\bc}{\begin{Corollary}} \newcommand{\ec}{\end{Corollary}}

\def\pd{\partial}

\begin{document}

\title{Saturation of the Quantum Null Energy Condition in Far-From-Equilibrium Systems}

\author{Christian Ecker}
\email{christian.ecker@tuwien.ac.at}
\affiliation{Institut f\"ur Theoretische Physik, Technische Universit\"at Wien \\
Wiedner Hauptstr.~8-10, A-1040 Vienna, Austria}

\author{Daniel Grumiller}
\email{grumil@hep.itp.tuwien.ac.at}
\affiliation{Institut f\"ur Theoretische Physik, Technische Universit\"at Wien \\
Wiedner Hauptstr.~8-10, A-1040 Vienna, Austria}

\author{Wilke van der Schee}
\email{wilke@mit.edu}
\affiliation{Center for Theoretical Physics, Massachusetts Institute of Technology,\\
77 Massachusetts Avenue, Cambridge, MA 02139, USA}
\affiliation{Institute for Theoretical Physics and Center for Extreme Matter and
Emergent Phenomena, Utrecht University, Leuvenlaan 4, 3584 CE Utrecht,
The Netherlands}

\author{Philipp Stanzer}
\email{pstanzer@hep.itp.tuwien.ac.at}
\affiliation{Institut f\"ur Theoretische Physik, Technische Universit\"at Wien \\
Wiedner Hauptstr.~8-10, A-1040 Vienna, Austria}

\begin{abstract}
The Quantum Null Energy Condition (QNEC) is a new
local energy condition that a general Quantum Field Theory (QFT) is believed to satisfy, relating the classical null energy condition (NEC) to the second functional derivative of the entanglement entropy in the corresponding null direction. 
We present the first series of explicit computations of QNEC in a strongly coupled QFT, using holography. We consider the vacuum, thermal equilibrium, a homogeneous far-from-equilibrium quench as well as a colliding system that violates NEC.
For vacuum and the thermal phase QNEC is always weaker than NEC. While for the homogeneous quench QNEC is satisfied with a finite gap, we find the interesting result that the colliding system can saturate QNEC, depending on the null direction.
\end{abstract}

\preprint{TUW--17--10, MIT--CTP/4954}

\keywords{Quantum Null Energy Condition, Gauge/Gravity Duality, Entanglement Entropy, Numerical Relativity, Non-Abelian Plasma Formation, Shockwave Collisions}
\pacs{21.65.Qr, 26.60.Kp, 11.25.Tq}

\maketitle

\section{Introduction}

Energy conditions rose to prominence in the 1960s as requisites for proofs of singularity theorems or Hawking's area theorem \cite{Hawking:1969sw, Hawking:1973uf}. While the specific energy condition needed depends on details of the particular theorem, all local classical ones are violated by quantum effects.
Even apparently feeble energy conditions such as NEC,
\be
\langle T_{kk} \rangle \equiv \langle T_{\mu\nu}k^\mu k^\nu \rangle \ge0\,,\qquad\forall\quad k^\mu k_\mu =0,
\label{eq:angelinajolie}
\ee
can be violated for stress tensors $T_{\mu\nu}$ in reasonable QFTs. Instead, QFTs typically obey non-local conditions such as the Averaged Null Energy Condition (ANEC, \cite{Klinkhammer:1991ki, Ford:1994bj}), which is the statement that negative energy density along a complete null geodesic is compensated by positive energy density (with ``quantum interest'' \cite{Ford:1999qv}).

These averaged energy conditions can sometimes be proven for QFTs (see \cite{Faulkner:2016mzt} for ANEC) and hence provide non-trivial consistency conditions for general QFTs. 
A better understanding of quantum energy conditions can then even lead to bounds on inflationary parameters, such as conjectured in \cite{Cordova:2017zej}.
Recently, inspired by singularity theorems of black hole dynamics a local quantum energy condition, QNEC, was proposed \cite{Bousso:2015mna}:
\be\label{eq:QNEC}
\langle T_{kk}\rangle\ge \frac{1}{2\pi \sqrt{h}}\,S''\,,\qquad\forall\quad k^\mu k_\mu =0\,.
\ee
Here $S''$ is the second functional derivative of entanglement entropy (EE) with respect to deformations of the entangling region along the null vector $k^\mu$ and $h$ denotes the determinant of the induced metric in the boundary of the entangling region (we set $\hbar=c=k_{\textrm{%
B}}=1$). %
Note that QNEC \eqref{eq:QNEC} is weaker (stronger) than NEC \eqref{eq:angelinajolie} if $S''$ is negative (positive).

Quantum energy conditions are particularly relevant for systems that violate the classical ones. A pertinent class of examples is provided by far from equilibrium strongly coupled quantum matter, which presents a challenge for most theoretical approaches. In this work we consider such examples.

While QNEC \eqref{eq:QNEC} is supposed to hold universally \cite{Balakrishnan:2017bjg}, most work so far \cite{Bousso:2015wca,Koeller:2015qmn,Akers:2017ttv,Fu:2017evt} focuses on holography \cite{Maldacena:1997re,Gubser:1998bc,Witten:1998qj}. This is because holography relates EE to simple geometrical entities in the dual gravitational bulk \cite{Ryu:2006bv,Hubeny:2007xt,Lewkowycz:2013nqa}, which would otherwise be notoriously hard to compute in the QFT itself.
An exception is \cite{Balakrishnan:2017bjg}, which generalized the proof of ANEC %
\cite{Faulkner:2016mzt} to prove QNEC for general QFTs.

QNEC is truly remarkable: it is the only known local energy condition that is supposed to hold in any relativistic QFT. Moreover, it relates a local quantity (the stress-tensor) to EE, which depends on the quantum state of the entangling region in question. We present several examples where indeed the inequality depends on the entangling region in a non-trivial way, but nevertheless QNEC is satisfied in all of them. 

Our Letter relies on previous work in numerical relativity
that determined the time-evolution of holographic entanglement entropy (HEE) \cite{AbajoArrastia:2010yt} and extracted features of interest for thermalization of anisotropic systems \cite{Ecker:2015kna} or holographic models of non-abelian plasma formation in heavy ion collisions \cite{Ecker:2016thn} based on a geometric setup that considers the collision of gravitational shockwaves \cite{Grumiller:2008va,Gubser:2008pc,Albacete:2008vs} numerically \cite{Chesler:2010bi,Casalderrey-Solana:2013aba,Chesler:2013lia}. This latter setup has the interesting property that for sufficiently localized shockwaves NEC \eqref{eq:angelinajolie} is violated \cite{Grumiller:2008va,Casalderrey-Solana:2013aba} with remarkable consequences for phenomenology, such as the absence of a local rest frame in far from equilibrium quantum matter \cite{Arnold:2014jva}.

The tools developed for calculating HEE can now be applied to evaluate QNEC %
numerically, and the present Letter reports the first such study. We consider physical systems of increasing complexity before finally addressing colliding gravitational shockwaves, where we discover a surprising saturation of the QNEC inequality \eqref{eq:QNEC}, depending on the null direction $k^\mu$ used therein. 

\section{Computing QNEC} 

We determine QNEC holographically by studying the gravitational dual, where EE of a region in the CFT can be computed using the Ryu-Takayanagi formula \cite{Ryu:2006bv,Hubeny:2007xt}.%
 \be\label{eq:RT}
S_\text{EE}=\frac{\mathcal{A}}{4G_N}=\frac{N_c^2}{2\pi} \mathcal{A}\equiv N_c^2\, \mathcal{S}_\text{EE}
\ee
Here $\mathcal{A}$ is the area of an extremal co-dimension 2 surface in the bulk which is homologous to the entangling region in the boundary and $G_N$ is Newton's constant. %
The prescription was proven in the static case \cite{Lewkowycz:2013nqa} and has survived many tests in dynamical situations \cite{Hubeny:2007xt,Callan:2012ip,Wall:2012uf}.

All our examples use metrics of the form
\be
\extd s^2=2\extd t\,(F \extd y-\extd z/z^2)-A\extd t^2+R^2\big(e^B\extd\mathbf{x}_{\perp}^{2}+e^{-2B}\extd y^2\big)
\label{eq:lalapetz}
\ee
where $A$, $B$, $F$ and $R$ can depend on boundary coordinates $t$, $y$ and the AdS radial coordinate $z$. These functions have normalizable modes $a_4$, $b_4$ and $f_4$ (with e.g.~$A=z^{-2}+a_4(t,\,y)z^2$), from which the projection of the stress tensor can be determined \cite{deHaro:2000vlm} as
\be
\label{eq:SEtensor}
\frac{1}{N_c^2} \langle T_{\mu\nu}k^\mu_\pm k^\nu_\pm\rangle \equiv \mathcal{T}_{\pm\pm} = \frac{1}{2\pi^2}(-a_4-2 b_4 \pm 2 f_4),
\ee
with null vectors $k_\pm^\mu = \delta_t^\mu\pm \delta_y^\mu$ at the boundary $z=0$.

In this work we consider entangling regions that are infinite strips along the perpendicular directions $x_\perp$ and hence are specified fully by their endpoints $S_\text{EE}(t_L, y_L; t_R, y_R)$ with a corresponding separation $L=y_R-y_L$. For these regions the extremal surface equation reduces to a geodesic equation in an auxiliary spacetime, which simplifies the computation considerably (see \cite{Ecker:2015kna,Ecker:2016thn} for a detailed description on the numerical procedure to find the relevant geodesics~\footnote{In this work we mostly work in a gauge where the horizon is at $z=1$ and the geodesics are parametrized by $t$ and the angle in the $y$-$z$-plane. To solve the equations we used a 5$^{\textrm{th}}$ order finite difference scheme with order 100 grid points. We verified our results with other gauges and numerical settings.}). The lengths of the geodesics then give the entropy density per transverse area. An important subtlety in computing \eqref{eq:RT} is its UV divergence. We regulate it by putting a cut-off at $z_\text{cut}=0.01$ and verifying that none of the physics presented in this Letter depends on the cut-off~\footnote{There is a subtlety in taking the functional derivative in Eq.~\eqref{eq:QNEC}, which can be UV-divergent if not taken in the right direction \cite{Bousso:2015mna}. We only deform the entangling region in the longitudinal direction, which then avoids this divergence.}.

After computing EE it is straightforward to evaluate QNEC \eqref{eq:QNEC} at some point $(t,\,y)$ for the null vectors $k_\pm^\mu$. This is done by computing $\pd_\lambda^2S_\text{EE}(t+\lambda,y\pm\lambda;\,t,y+L)$ at $\lambda=0$, which yields $S''/\sqrt{h}$ in Eq.~\eqref{eq:QNEC}~\footnote{In this Letter all set-ups are invariant under $y\to-y$ so without loss of generality we only vary the left point of the entangling region.}.

\begin{figure*}[ht!]
\center
\includegraphics[width=0.7\textwidth]{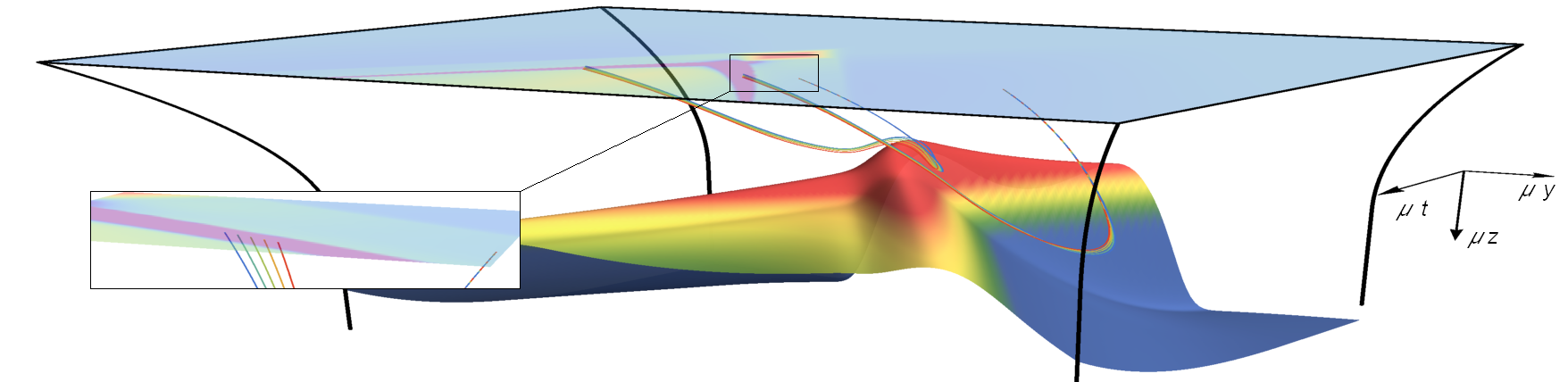}
$\;\,$ \includegraphics[width=0.27\linewidth]{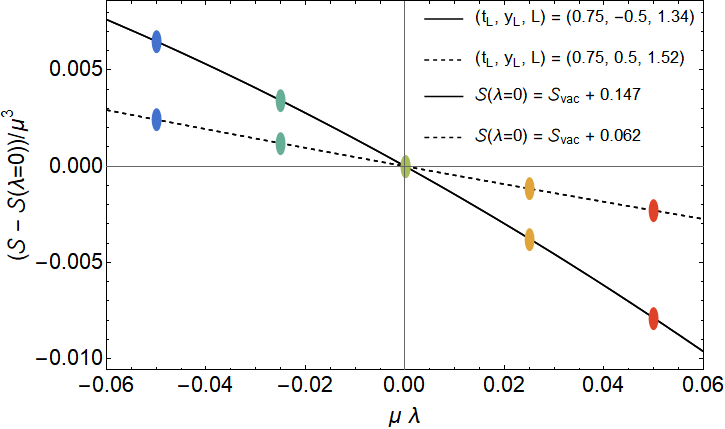}
\caption{
Two families of extremal surfaces at representative locations $(t_L,\,y_L)=(0.75,\,\pm 0.5)$ for the shockwave geometry. The families correspond to a null variation at a point where the classical NEC is violated (purple region at $z=0$ or black region in left Fig.~\ref{fig:ShockWaves}). The family starting at $y=-0.5$ hovers just above the apparent horizon (colored surface) and hence has larger entropy, as well as more negative $\mathcal{S}''_\pm$ (right Fig., see also Fig.~\ref{fig:QNECshocks2}).
\label{fig:QNECshocks}}
\end{figure*}

It is instructive to examine QNEC from a near-boundary perspective, where it is possible to prove QNEC \cite{Koeller:2015qmn}. Close to the boundary point $(t_L,\,y_L)$ an extremal surface is given by
$t(z)=t_L+\lambda-z+t_4(\lambda)z^4+a_4z^5/5+O(z^6)$,
$y(z)=y_L\pm\lambda-z+y_4(\lambda)z^4+f_4z^5/5+O(z^6)$,
where $t_4$ and $y_4$ also depend on $(t_R,\,y_R)$ and are undetermined in a near-boundary analysis. Extremal surfaces are stationary under perturbations, so variations of extremal surfaces only yield boundary terms. A simple geometric argument then gives
$\partial_\lambda\mathcal{A}=-4t_4(\lambda)\pm4y_4(\lambda)$,
which leads to the second variation $S''_\pm = (\pm 4 \partial_\lambda y_4 - 4 \partial_\lambda t_4) / (4 G_N)$.

Since we perturb in a null direction the leading contribution to the distance between the two extremal surfaces separated by $\lambda$ vanishes. We have two subleading contributions, coming from the subleading terms in the extremal surface and metric expansions respectively:
\begin{align}
\Delta s^2 &= |x^\mu(t_L, y_L, z) - x^\mu(t_L + \lambda, y_L + \lambda, z)|  \nonumber \\
&=z^2 \lambda^2 (-2 b_4 \pm 2 f_4 - a_4 \mp 2 \partial_\lambda y_4 + 2 \partial_\lambda t_4) 
\label{eq:ds2proof}
\end{align}
Assuming the classical NEC in the bulk spacetime and using that the deformation along $\lambda$ is null, it can be shown \cite{Wall:2012uf} that the distance between the surfaces has to be spacelike, i.e., $\Delta s^2 \geq 0$, also called `entanglement nesting property'. This condition reduces precisely to QNEC in \eqref{eq:QNEC}.
Equation \eqref{eq:ds2proof} is useful for us, not only to illustrate why in holography we expect QNEC to be valid, but also to independently verify QNEC from a bulk perspective.
This is done by explicitly computing the distance between two nearby extremal surfaces and comparing this with QNEC determined as described next.

To evaluate QNEC in practise we evaluate the second derivative by computing $\mathcal{S}_{\textrm{\tiny{EE}}}$ for five equidistant values of $\lambda$ between $-0.05$ and $0.05$. We then obtain four estimates of $\mathcal{S}''_\pm$ by generating a quadratic fit through all five points, the first three points, the middle three points and the last three points, thereby both obtaining a mean estimate as well as a numerical error. 

Figure~\ref{fig:QNECshocks} shows an example of a family of surfaces for $k_+^\mu$ at $t_L=t_R=0.75$, $y=0.5$ and $L=1.0$, including the apparent horizon of the shockwave collisions and the (violation of) NEC on the boundary. On the right we display EE of the five surfaces, having their vacuum contribution subtracted.

To obtain the full QNEC result it is necessary to add the vacuum contribution again. This is straightforward, since for a strip the vacuum EE per transverse area is known analytically \cite{Ryu:2006bv},
\be\label{eq:Svac}
\mathcal{S}_\text{EE} =\frac{1}{2\pi} \bigg( \frac{1}{z_\text{cut}^2} - \frac{1}{2c_0^3l^2} \bigg)\qquad c_0=\frac{3\Gamma[1/3]^3}{2^{1/3}(2\pi)^2}
\ee
where $l=\sqrt{(L\pm\lambda)^2-\lambda^2}$ is the proper length of the (boosted) strip. Taking the second derivative with respect to $\lambda$ at $\lambda = 0$ gives
\be\label{eq:Sppvac}
\frac{1}{2\pi}\,\mathcal{S}''_\pm = -\frac{1}{\pi^2c_0^3L^4}\approx -\frac{0.06498}{L^4} \,.
\ee
From Eq.~\eqref{eq:Sppvac} it is clear that the CFT vacuum satisfies QNEC in a trivial way, especially for small $L$, while it saturates QNEC in the limit $L\to\infty$.

\section{Results}

\paragraph{Thermal plasma - } We first consider a homogeneous thermal equilibrium state with dual description in terms of the AdS$_5$ Schwarzschild black brane that has $A=1/z^2-({\pi}T)^4z^2$, $R=1/z$ and $B=F=0$, where the energy density is related to the temperature by $T^0_0=3N_c^2\pi^2T^4/8$. The null projections of the energy momentum tensor, $\mathcal{T}_{\pm\pm}$, are the same for both lightlike directions due to parity symmetry,
\be
\frac{1}{N_c^2} \langle T_{\mu\nu}k^\mu_\pm k^\nu_\pm\rangle \equiv \mathcal{T}_{\pm\pm} = \frac{\pi^2}{2}\,T^4\approx 0.0507 \,\pi^4 T^4\,.
\ee
In this case $\mathcal{S}''_+\!=\!\mathcal{S}''_-$, which can be understood by realizing that the plasma is time-reversal invariant. That means we can invert the $k_t$ component and invariance of the second derivative under $k_\mu\to-k_\mu$ yields the identity.

\begin{figure}[ht]
\center
\includegraphics[width=0.52\linewidth]{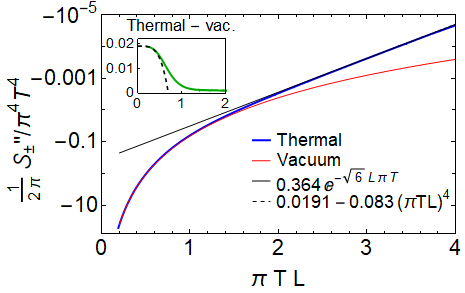}
\includegraphics[width=0.45\linewidth]{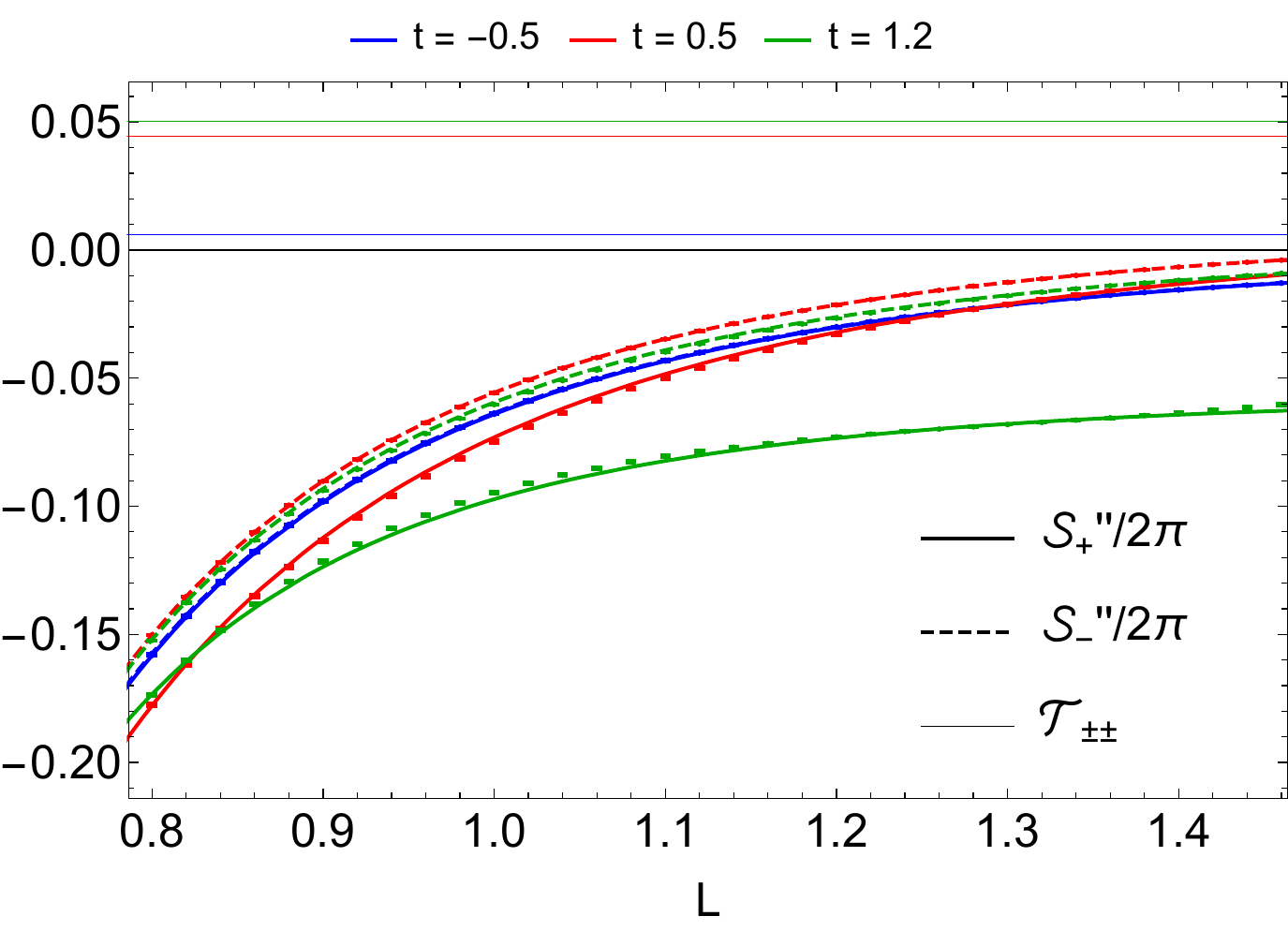}
\caption{
Left: $\mathcal{S}''_\pm$ for thermal state as function of strip length (blue). For small $L$ the curve follows the vacuum result (Eq.~\eqref{eq:Sppvac}, red) whereas for large length $\mathcal{S}''_\pm$ approaches zero exponentially (black). %
Since $\mathcal{S}''_\pm<0$ and $\mathcal{T}_{\pm\pm}>0$ QNEC is obviously satisfied.
Right: $\mathcal{S}''_\pm$ at three different times as function of separation $L$ together with constant $\mathcal{T}_{\pm\pm}$ [for the quenched geometry \eqref{eq:lalapetz} with \eqref{eq:MVaidya} and $B\!=\!F\!=\!0$].
}
\label{fig:QNECblackbrane}
\end{figure}

In Fig.~\ref{fig:QNECblackbrane} we show that at small length $\mathcal{S}''_\pm$ approaches the vacuum result, while for large $L$ it approaches zero from below exponentially fast. Since $\mathcal{T}_{\pm\pm}$ is positive we see that QNEC is easily satisfied for all lengths and  never saturates. Analytic calculations in the Supplemental Material confirm our numerical results at small and large $L$.

\paragraph{Far-from-equilibrium quench - } Now we consider a quenched far-from-equilibrium system where a homogeneous shell of null dust is injected in the gravitational dual \cite{AbajoArrastia:2010yt}, leading to the AdS$_5$ Vaidya spacetime
\begin{align}
A=z^{-2} - M(t) z^2\,,\qquad M(t)\equiv\tfrac{1}{2}\left(1+\tanh(2 t)\right)\,.
\label{eq:MVaidya}
\end{align}
Equation~\eqref{eq:MVaidya} realizes a homogeneous quench of the vacuum at $t\!=\!-\infty$ to a thermal state with $T\!=\!\frac{1}{\pi}$ at $t\!=\!\infty$. The corresponding projection of the energy momentum tensor is time dependent, with $\mathcal{T}_{\pm\pm}=\frac{1}{2\pi^2}M(t)$. The Vaidya geometry is not invariant under time inversion, so $\mathcal{S}''_\pm$ are distinct from each other.
\begin{figure}[ht]
\center
\includegraphics[width=0.47\linewidth]{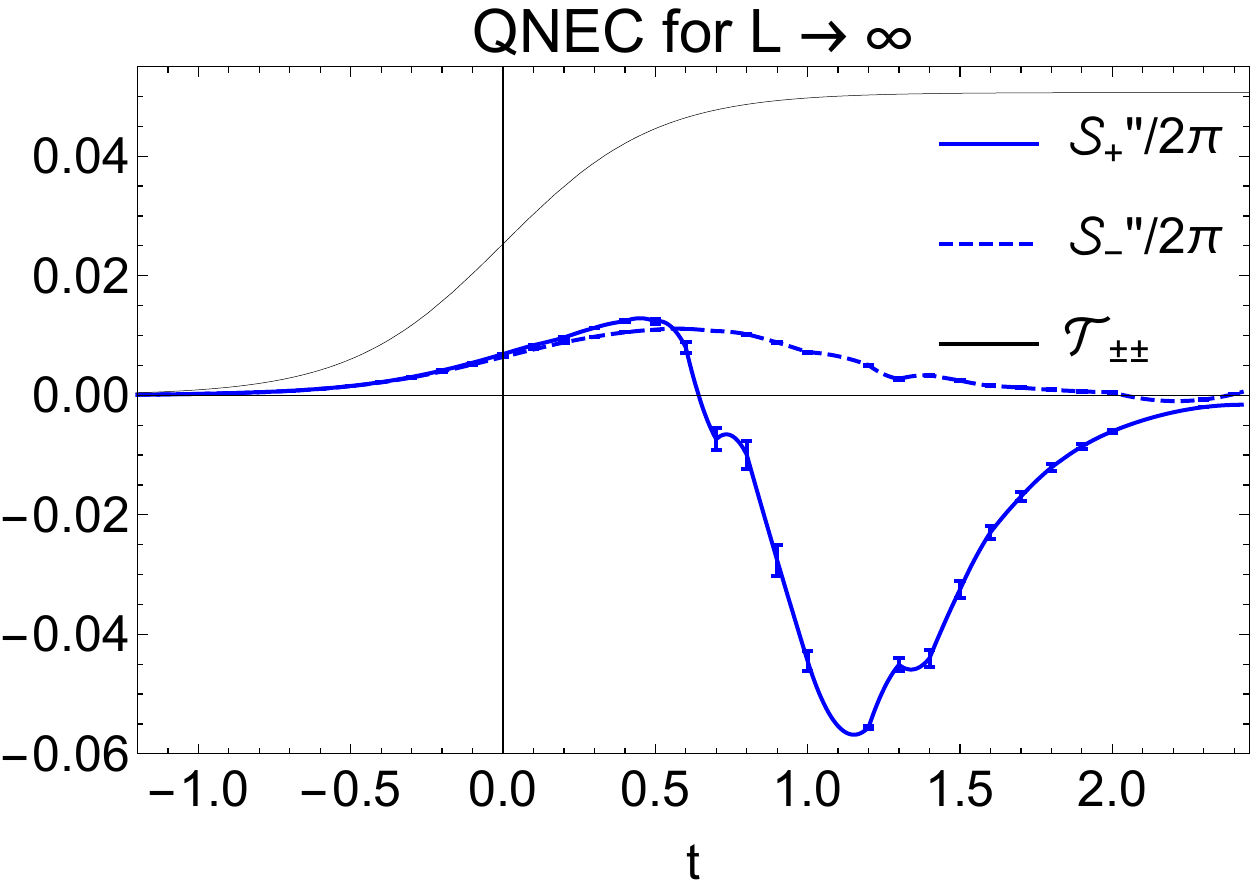}
\includegraphics[width=0.47\linewidth]{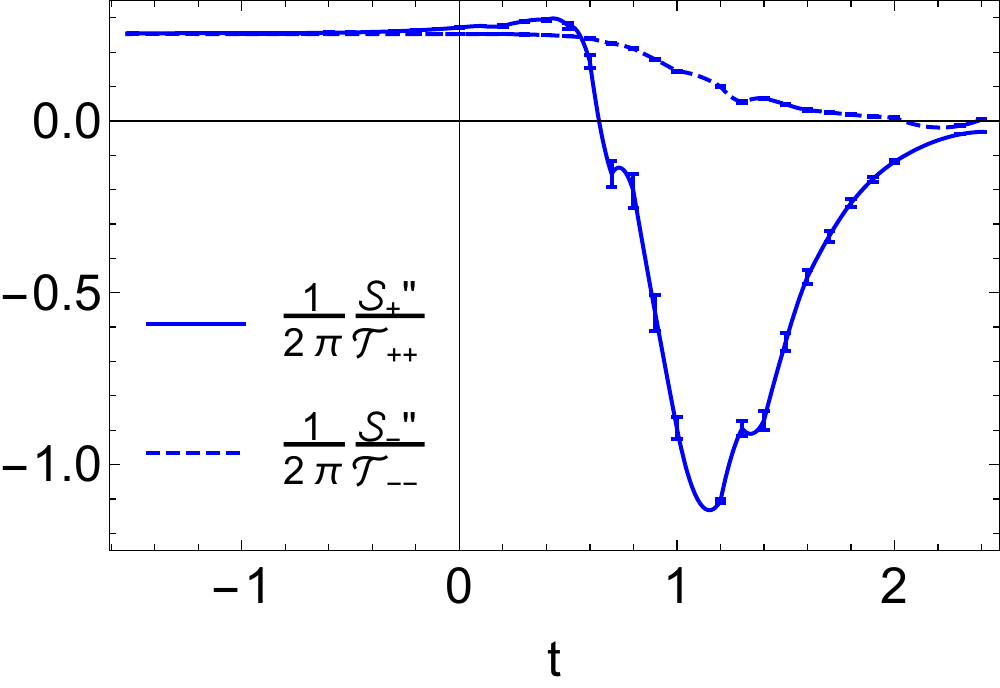}
\caption{Left: long length limit of right Fig.~\ref{fig:QNECblackbrane}.  Right: ratio of the two sides of the QNEC inequality \eqref{eq:QNEC}. Curiously the ratio asymptotes to $0.25$ at early times and never goes above $0.30$. QNEC is still non-trivial for a time of order $1/({\pi}T)$ after the geometry has already settled down.
}
\label{fig:QNECvaidya}
\end{figure}

\newcommand{\lcc}{x}

\begin{figure*}[ht]
\center
\includegraphics[width=0.31\textwidth]{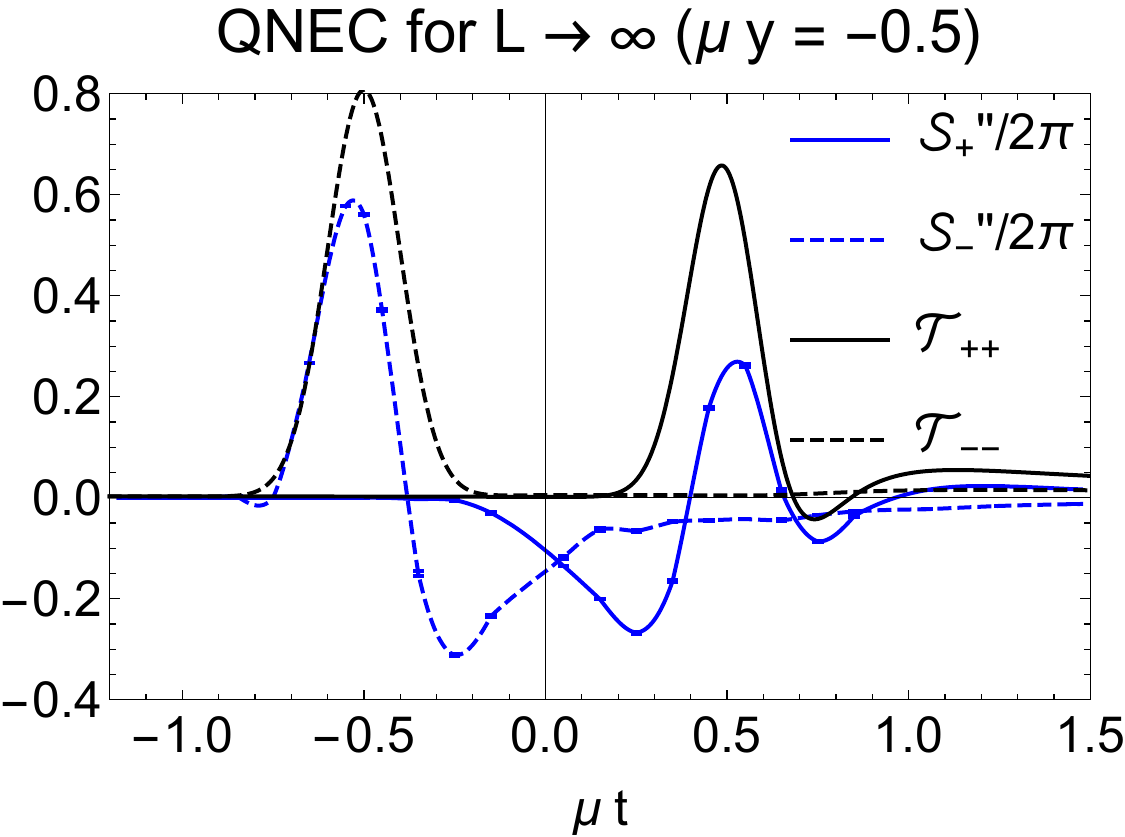}
\includegraphics[width=0.31\textwidth]{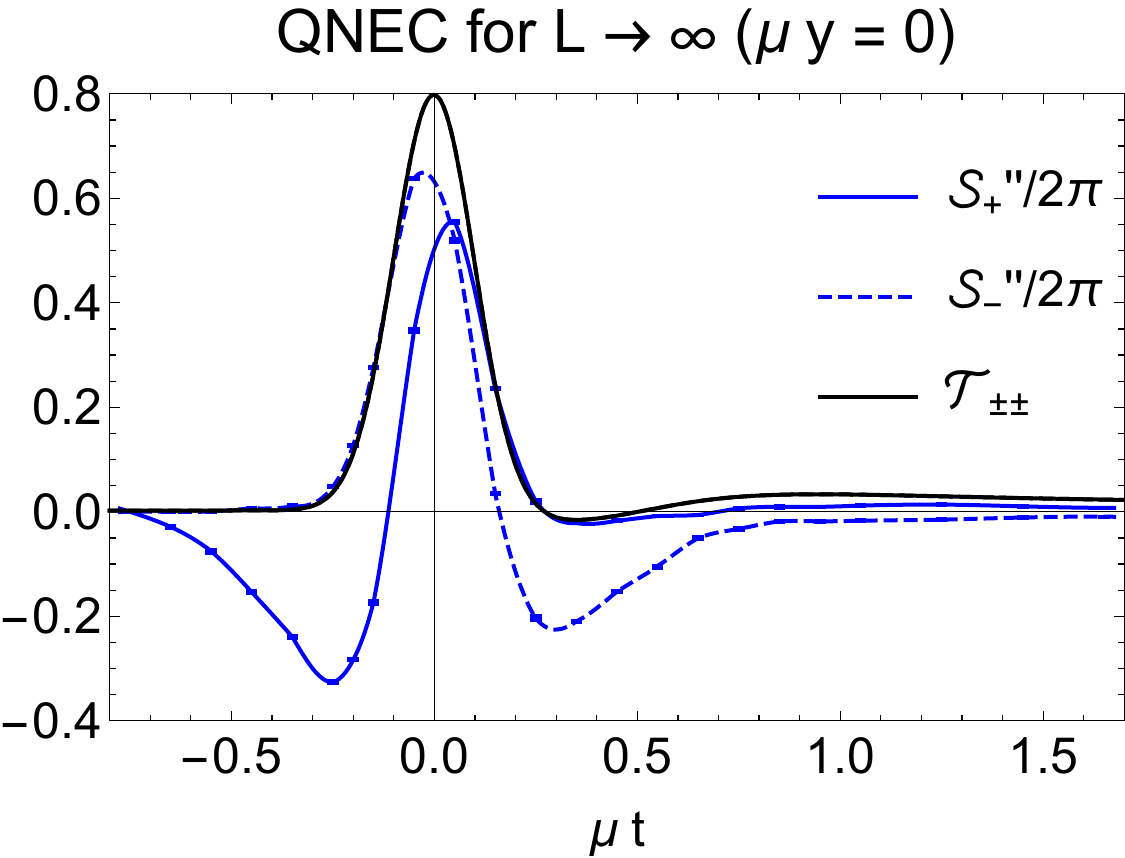}
\includegraphics[width=0.31\textwidth]{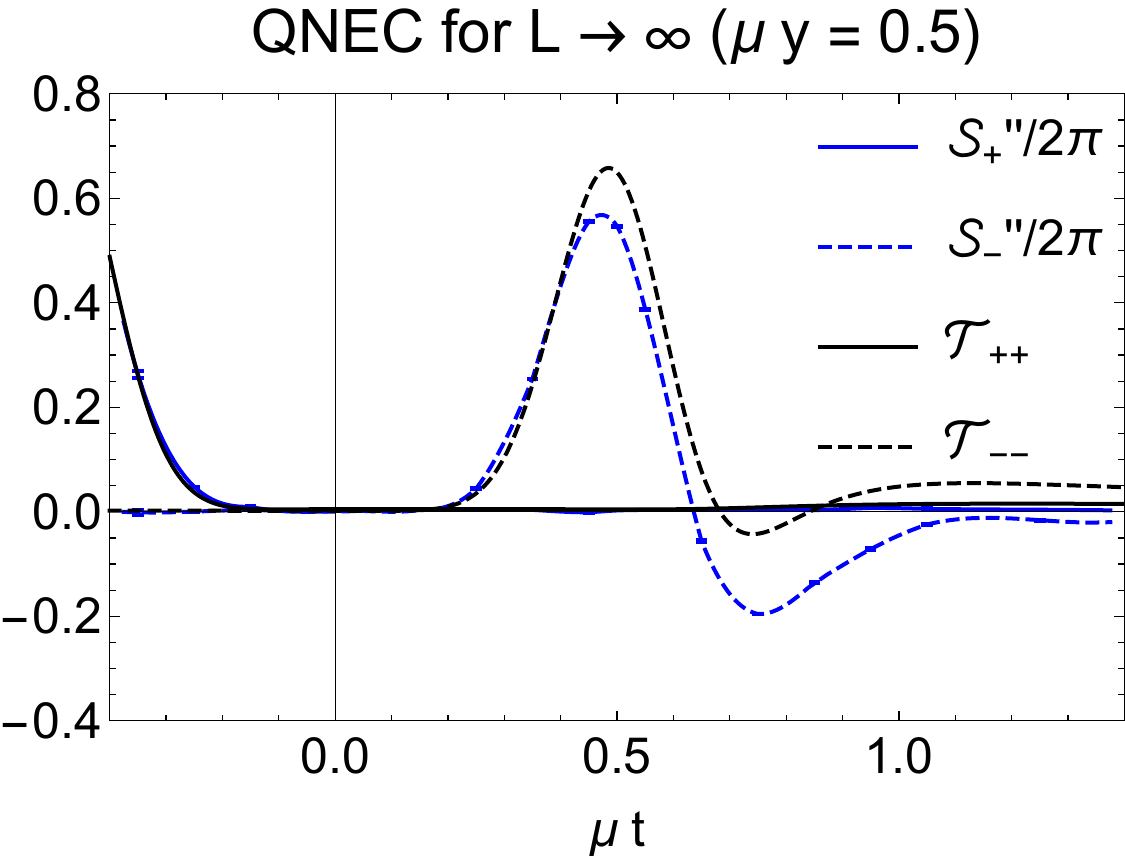}
\caption{Large $L$ limit of QNEC as fuction of time for $y=-0.5$ (left), $y=0$ (middle) and $y=0.5$ (right). Strikingly, depending on the direction of $k_\mu$ all cases show a saturation of QNEC in the far-from-equilibrium regime, where in the center case first the $k_-$ direction saturates, after which it transitions to the $k_+$ direction, which saturates when NEC is violated ($\mathcal{T}_{\pm\pm}<0$).
}
\label{fig:QNECshocks2}
\end{figure*}

In the right Fig.~\ref{fig:QNECblackbrane} we show $\mathcal{S}''_\pm$ versus the length of the strip at three different times.
The plot includes fits of the form
$\mathcal{S}''_\pm(L)=Q_\pm+a_{\pm}L^{-b_\pm}$,
obtained from the data points between $L=0.6$ and $L=1.5$ with weights proportional to the inverse error squared (for instance, for $t=0.5$ we obtain $Q_+=0.0123 \pm 0.0004$, $Q_-=0.01098 \pm 0.00001$, $a_+=-0.0854\pm 0.0006$, $a_-=-0.0666\pm 0.00002$, $b_+=3.59\pm 0.01$, and $b_-=3.962\pm 0.0007$).
For all cases presented this gives an accurate estimate for the large length value $Q_\pm$ and its numerical error from the least-square fit, which we show in Fig.~\ref{fig:QNECvaidya}. Even though the geometry is only slightly perturbed at early times, we curiously see that the ratio of $\mathcal{S}''_\pm/(2\pi)$ versus $\mathcal{T}_{\pm\pm}$ reaches a constant value of about 0.25. We also see that QNEC settles down to its thermal value later than the stress-tensor itself. 
This setting is the first case where QNEC is stronger than NEC, i.e.~we find cases with $\mathcal{S}''_\pm > 0$. Nevertheless, QNEC never saturates, with  $\mathcal{S}''_\pm/(2\pi)$ only attaining at most 30\% of $\mathcal{T}_{\pm\pm}$ in our example.

\begin{figure}[htb]
\center
\includegraphics[width=0.42\linewidth]{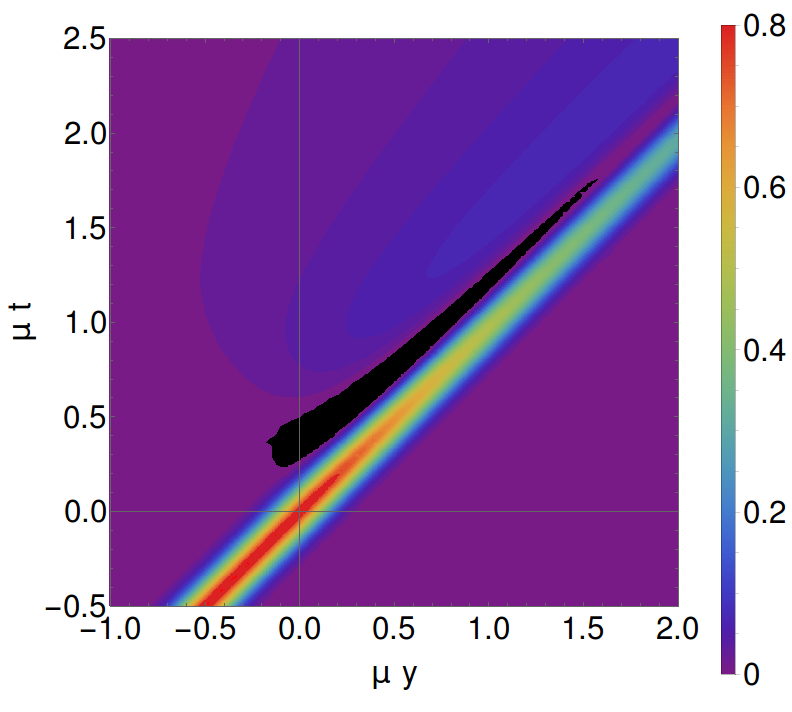}\includegraphics[width=0.03\linewidth]{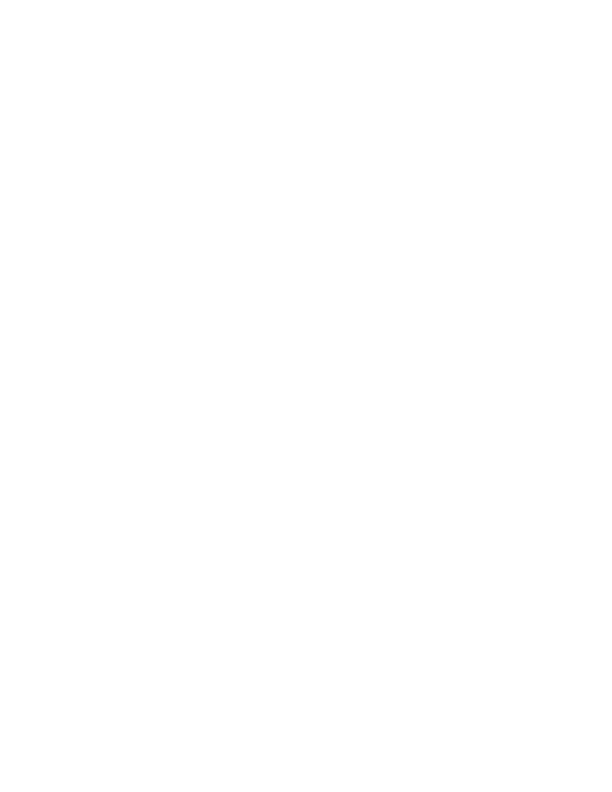}\includegraphics[width=0.55\linewidth]{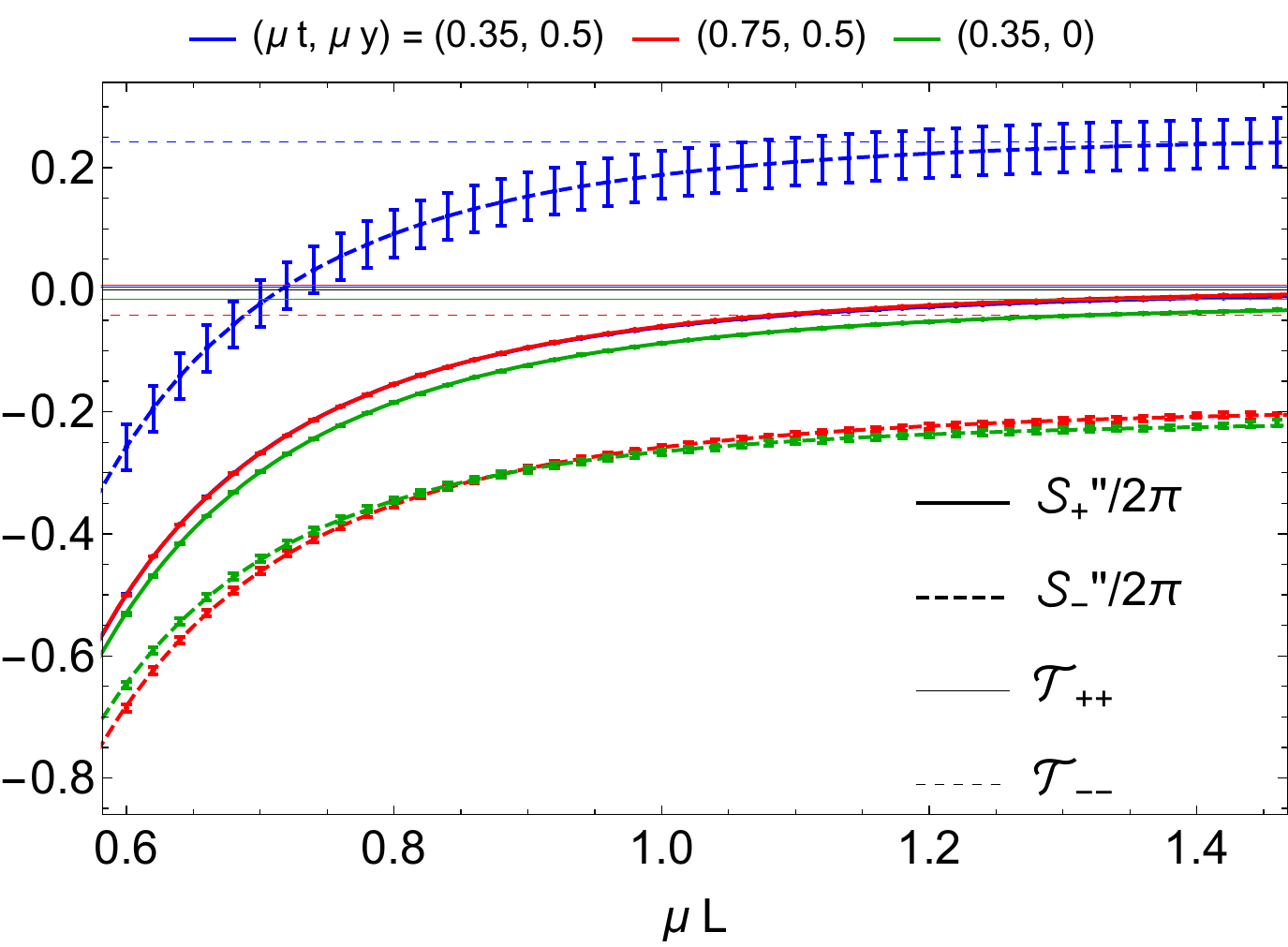}
\caption{%
Left: Contour plot of ${\cal{T}}_{--}$ with NEC violation in black region. 
Right: QNEC terms as function of $L$ for three representative points in shockwave geometry (see Figs.~\ref{fig:QNECshocks},~\ref{fig:QNECblackbrane}). Dashed blue saturates QNEC, even though NEC is positive. Dashed red violates NEC, but $\mathcal{S}''_{-}$ is even smaller and no saturation occurs in $-$ direction, while it occurs in $+$ direction.}
\label{fig:ShockWaves}
\end{figure}

\paragraph{Shockwave collision - } The richest example presented here analyzes QNEC for the CFT state dual to colliding gravitational shockwaves. This in particular leads to regions where the ordinary NEC is violated \cite{Arnold:2014jva} and hence gives a perfect setting to examine QNEC. Colliding shockwaves are dual to planar sheets of energy moving at the speed of light and fully characterized by their only non-zero component of the boundary stress-energy tensor $\mathcal{T}_{\pm\pm}=1/2\pi^2h_\pm(\lcc_\pm)$, with $\lcc_\pm=t{\pm}y$, where $h_\pm(\lcc_\pm)=\mu^3\exp[-\lcc_\pm^2/2w^2]/\sqrt{2{\pi}w^2}$ and ${\mu}w=0.1$. 
We determined the functions $A$, $B$, $F$ and $R$ in the metric \eqref{eq:lalapetz} numerically in previous work \cite{Casalderrey-Solana:2013aba} and use these results here as input for our evaluation of HEE and QNEC.

Figure~\ref{fig:QNECshocks} shows the bulk shockwave evolution, whereby the colors at $z=0$ represent (the violation of) NEC (see also \cite{Arnold:2014jva} and the left Fig.~\ref{fig:ShockWaves}). The right Fig.~\ref{fig:ShockWaves} shows analogous $\mathcal{S}''_\pm$ versus $L$ plots at three representative points, noting that $\mathcal{T}_{++}$ differs from $\mathcal{T}_{--}$ at $y \neq 0$. The red curve is at the location where NEC is significantly violated, with $\mathcal{T}_{--}\!=\!-0.04\mu^4$, while QNEC is satisfied, with $\mathcal{S}''_{-}/(2\pi)$ asymptoting to $-0.19\mu^4$. For $k_\mu\!=\!k_+$ NEC is satisfied, but QNEC is saturated, with $\mathcal{T}_{++}\!=\!\mathcal{S}''_{+}/(2\pi)\!=\!0.01\mu^4$.

Figure~\ref{fig:QNECshocks2} shows the asymptotic behavior of QNEC for $\mu y=-0.5,\,0.0\text{\;and\;}0.5$ [recall that $\pm0.5$ are distinct from each other due to our choice of varying the left point of the strip in Eq.~\eqref{eq:QNEC}].

Strikingly, at $y=0$ we find QNEC saturation in the far-from-equilibrium regime for $k_-$ at negative times, which transitions to saturation for $k_+$ at positive times. During the hydrodynamic phase at ${\mu}t>0.8$ there is no saturation. For $y=0.5$ we have the non-trivial result that QNEC is saturated for both $k_-$ and $k_+$ as the outgoing shock passes around ${\mu}t=0.3-0.5$. Lastly, for $y=-0.5$ the entangling region encompasses most of the collision region and we do not find saturation for $t>0$.

\section{Discussion} 

Our main result is the saturation of QNEC in far-from-equilibrium regions created during the shockwave collisions. This saturation is non-trivial and not seen in other systems we studied. For vacuum and thermal states QNEC is weaker than NEC, since $\mathcal{S}''$ is always negative. For a homogeneous quench QNEC is stronger than NEC, but the ratio of both sides of the inequality never exceeds $0.3$. In the shockwave collisions QNEC is never saturated in the hydrodynamic regime, but it is saturated in the far-from-equilibrium region, regardless of whether NEC is valid. Reference \cite{Koeller:2017njr} (see also \cite{Casini:2017roe}) conjectures that saturation of QNEC can lead to a simplified expression for (part of) the modular Hamiltonian of a half-space in vacuum.

Even in vacuum QNEC is non-trivial, as for our strip the EE term scales as $\mathcal{S}_\pm''\propto-1/L^4$, which has a UV divergence as $L\to0$. This makes the inequality trivially satisfied in the small length limit, and it is hence an interesting question whether QNEC also holds if one looks at a more physical quantity, such as the vacuum-subtracted EE. None of the proofs of QNEC apply for that case, but for all points where we checked QNEC we found that this stronger condition also holds. 

QNEC is a remarkable quantum inequality, and examples such as the ones studied in the Letter will help to further explore its more general implications as well as applications such as holographic descriptions of strongly coupled quantum matter.

\acknowledgments

\paragraph{Acknowledgments - } 
We thank Frederic Br\"unner, Matthew Headrick, Veronika Hubeny, Ville Ker\"anen, Aitor Lewkowycz, David Mateos, Ayan Mukhopadhyay, David M\"uller, Florian Preis, Anton Rebhan, Stefan Stricker, and Aron Wall for useful discussions. This work is supported by the Austrian Science Fund (FWF), projects P27182-N27, P28751-N27 and DKW1252-N27.  WS is supported by the U.S. Department of Energy under grant Contract Number DE-SC0011090 and by VENI grant 680-47-458 from the Netherlands Organisation for Scientific Research (NWO).

\section{SUPPLEMENTAL MATERIAL}

\setcounter{equation}{0}
\renewcommand{\theequation}{S\arabic{equation}}

\subsection{QNEC for AdS\texorpdfstring{$_5$}{5} Schwarzschild black brane}

\paragraph{Preliminaries - } HEE for the AdS$_d$ Schwarzschild black brane was considered by Fischler and Kundu who gave infinite series representations in terms of ratios of $\Gamma$-functions \cite{Fischler:2012ca} and more recently by Erdmenger and Miekley \cite{Erdmenger:2017pfh} who expressed their results in closed form in terms of Meijer G-functions. For QNEC it is necessary to compute non-equal time HEE, which is not straightforward using these methods. We use a more pedestrian approach that allows straightforward generalization from HEE to QNEC as well as fast and precise numerical evaluation of QNEC at small and large separations. For sake of specificity we focus on $d=5$, but our methods and results can be generalized easily to arbitrary dimensions. In this way we shall recover the vacuum result for HEE \eqref{eq:Svac} and QNEC \eqref{eq:Sppvac} as well as the corresponding thermal results in the main text, see the left Fig.~\ref{fig:QNECblackbrane}.

\newcommand{\hor}{(\pi T)}

\paragraph{Geometry - } The AdS$_5$ Schwarzschild black brane metric is given by
\eq{
\extd s^2 = \frac{1}{z^2}\,\Big(-f(z)\,\extd t^2 + \frac{\extd z^2}{f(z)} + \extd y^2 + \extd x_1^2+\extd x_2^2\Big)
}{eq:s1}
with
\eq{
f(z)= 1 - (\pi T)^4 z^4  
}{eq:s2}
where $T$ is the Hawking temperature in the same units as in the main text.

\paragraph{Area functional - } For a strip the minimal area per transverse density functional reads
\eq{
{\cal A} = %
\int\limits_0^{\frac{\ell+\lambda}{2}-\omega}\!\!\!\!\extd y\, L(z,\,\dot z,\,\dot t)
}{eq:s3}
with Lagrangian
\eq{
L(z,\,\dot z,\,\dot t) = \frac{2}{z^3}\,\sqrt{1+\frac{\dot z^2}{f(z)}-\dot t^2f(z)} 
}{eq:s4}
where the dimensionful quantity $\ell$ is the width of the strip in $y$-direction before deformation and $\lambda$ parametrizes the null deformation of the boundary interval with boundary points $(t_\pm,\,y_\pm)=(\pm\lambda/2,\,\pm(\ell+\lambda)/2)$. This means that for $\lambda=0$ we shall recover the HEE results for a strip of width $\ell$ centered around $y=0$ at the constant time-slice $t=0$. Moreover, $\omega$ denotes the cut-off on the holographic coordinate, such that $z(\ell/2-\omega)=z_\text{cut}\ll 1$, dots denote derivatives with respect to $y$ and the overall factor $2$ in \eqref{eq:s4} comes from the fact that we have two equally big contributions to the area by integrating $y$ from the midpoint $y=0$ to either of the endpoints $y_\pm=\pm(\ell+\lambda)/2$. 

\paragraph{Noether charges - } Since the functional \eqref{eq:s3} respects translation invariance, $y\to y+y_0$, there is an associated Noether charge yielding a first integral,
\eq{
Q_1 = L - \dot z \,\frac{\partial L}{\partial\dot z} = \frac{2}{z^3\sqrt{1+\dot z^2/f(z)-\dot t^2 f(z)}} =: \frac{2}{z_\ast^3 N_\ast}
}{eq:s5}
with 
$N_\ast=\sqrt{1-(\dot t^2 f)|_{z=z_\ast}} = \sqrt{1-\Lambda^2/f(z_\ast)}$
chosen such that at $z(y\to 0)=z_\ast$ we are at the tip of the extremal surface, $\dot z=0$. The constant $\Lambda=(\dot t f)|_{z=z_\ast}$ was introduced in anticipation of \eqref{eq:s25} below.

There is a second Noether charge following from $\partial_y (\partial L/\partial\dot t)=0$, yielding a constant of motion $\Lambda$.
\eq{
Q_2 = \dot t f(z) =: \Lambda 
}{eq:s25}
Combining the two Noether charges $Q_{1,2}$ establishes an expression for $\dot z$.
\eq{
\dot z = - \sqrt{\big(N_\ast^2 z_\ast^6/z^6-1\big)\,f(z) + \Lambda^2}
}{eq:s34}

The values of the two Noether charges are fixed by the interval parameters $\ell$ and $\lambda$. Integrating \eqref{eq:s34} from the tip of the surface $z=z_\ast$ to the boundary $z=0$ and introducing the dimensionless variable $x=z/z_\ast$ yields
\eq{
\frac{\ell+\lambda}{2} = z_\ast\,\int\limits_0^1\extd x\,\frac{x^3}{R(x)} 
}{eq:s38}
with
$R(x):=\sqrt{(N_\ast^2-x^6)(1-(\pi Tz_\ast x)^4)+ \Lambda^2x^6}$.
Similarly, integrating $\dot t$ from $t=0$ to $t=\lambda/2$ (which again can be converted into a $z$-integration from the tip of the surface $z=z_\ast$ to the boundary $z=0$) yields
\eq{
\frac{\lambda}{2}=\Lambda\, z_\ast\int\limits_0^1\extd x\,\frac{x^3}{f(xz_\ast)\,R(x)} \,.
}{eq:s100}
For small $\ell$ it is useful to determine $\Lambda$ instead from
\eq{
\Lambda = \frac{\lambda}{\ell+\lambda+2z_\ast I_{\Delta}}
}{eq:s101}
with 
\eq{
I_{\Delta} = \int\limits_0^1\extd x\,\frac{x^3}{R(x)}\bigg(\frac{1}{f(xz_\ast)}-1\bigg)\,.
}{eq:s102}
For QNEC we need to expand to order ${\cal O}(\lambda^2)$ but not higher, which means that in \eqref{eq:s101} we need to take into account only terms in $I_{\Delta}$ of order unity or linear in $\Lambda$, but no higher powers of $\Lambda$.

\paragraph{Area as integral - } Inserting the first integrals \eqref{eq:s34}, \eqref{eq:s25} into the area functional \eqref{eq:s3} with \eqref{eq:s4} and \eqref{eq:s2} and expanding in powers of the cutoff $z_{\textrm{cut}}$ yields
\eq{
{\cal A} = \frac{1}{z_\text{cut}^2} + \frac{2}{z_\ast^2}\,\big(I_{\cal A}^{\lambda}-\tfrac12\big) + {\cal O}(z_{\textrm{cut}}^2)
}{eq:s36}
with the finite contribution
\eq{
I_{\cal A}^{\lambda} = \int\limits_0^1\extd x\,\frac{1}{x^3}\,\bigg(\frac{N_\ast}{R(x)}-1\bigg)\,.
}{eq:s37}

The remaining task in order to get the area as function of the dimensionless product of temperature and strip width, $T\ell$, is to evaluate the integrals \eqref{eq:s37}, \eqref{eq:s102} and \eqref{eq:s38}. We consider first the limit of small widths, $T\ell \ll 1$, and then of large widths, $T\ell \gg 1$. These results will allow comparison with the numerical fits in the main text and in Fig.~\ref{fig:QNECblackbrane}.

\paragraph{Small width expansion - } We start with the small width expansion $T\ell\ll 1$. Note that we have the chain of inequalities $0<\lambda/\ell\ll T\ell\ll 1$. As we shall see, all our results are expressed succinctly in powers of a single transcendental number, 
\eq{
 c_0 = \frac{3\Gamma[1/3]^3}{2^{1/3}(2\pi)^2} \approx 1.159595
}{eq:c0}
which was already introduced in the main text \eqref{eq:Svac}. Perturbative evaluation of the integral \eqref{eq:s102} together with \eqref{eq:s101} yields
\begin{multline}
\Lambda = \tfrac{\lambda}{\ell+\lambda}  - (\pi Tz_\ast)^4\,\tfrac{4\pi c_0\lambda z_\ast}{15 \sqrt{3}(\ell+\lambda)^2} + (\pi Tz_\ast)^8\, \Big(\tfrac{16\pi^2 c_0^2\lambda z_\ast^2}{675(\ell+\lambda)^3} \\
-\tfrac{2\lambda z_\ast}{3(\ell+\lambda)^2}\Big) + {\cal O}((T z_\ast)^{12}) + {\cal O}(\lambda^3/\ell^3)\,.
\label{eq:s105}
\end{multline}
Similarly, evaluation of the integral \eqref{eq:s38} establishes a series expansion for $z_\ast$,
\begin{multline}
\frac{z_\ast}{c_0\ell} = 1 + (\pi T\ell)^4\,\tfrac{2\pi c_0^6}{15\sqrt{3}} + (\pi T\ell)^8\, \big(\tfrac{4\pi^2 c_0^{12}}{135}-\tfrac{c_0^9}{6}\big)\\
+\tfrac{\lambda}{\ell}\, \Big(1 - (\pi T\ell)^4\,\tfrac{2\pi c_0^6}{3\sqrt{3}} + (\pi T\ell)^8\, \big(\tfrac{4\pi^2c_0^{12}}{15}-\tfrac{3c_0^9}{2}\big)\Big)\\
+\tfrac{\lambda^2}{\ell^2}\, \Big(-\tfrac12 + (\pi T\ell)^4\,\big(\tfrac{c_0^4}{6} - \tfrac{49\pi c_0^6}{45\sqrt{3}}\big) + (\pi T\ell)^8\, \big(\tfrac{c_0^8}{6} - \tfrac{71 c_0^9}{12} \\
- \tfrac{c_0^{10} \pi}{5 \sqrt{3}} + \tfrac{2074 c_0^{12}\pi^2}{2025} \big) \Big)
+{\cal O}((T\ell)^{12})+{\cal O}(\lambda^3/\ell^3) 
\label{eq:s106}
\end{multline}
where we additionally expanded in powers of the dimensionless small parameter $\lambda/\ell$, keeping only the powers needed to determine QNEC. Finally, the area integral \eqref{eq:s37}, together with the other results above, leads to an expression for the area \eqref{eq:s36}
\begin{multline}
{\cal A} = \tfrac{1}{z_{\textrm{cut}}^2} - \tfrac{1}{2c_0^3\ell^2} + \hor^4\ell^2\,\tfrac{\pi c_0^3}{5\sqrt{3}} + \hor^8\ell^6\,\big(\tfrac{c_0^6}{12} - \tfrac{2 c_0^9 \pi^2}{225}\big) \\
+\tfrac{\lambda}{\ell}\,\Big(\tfrac{1}{c_0^3\ell^2} + \hor^4\ell^2\,\tfrac{2\pi c_0^3}{5\sqrt{3}} + \hor^8\ell^6 \big(\tfrac{c_0^6}{2} - \tfrac{4 c_0^9 \pi^2}{75}\big)\Big) \\
+\tfrac{\lambda^2}{\ell^2}\,\Big(-\tfrac{2}{c_0^3\ell^2} + \hor^4\ell^2\,\tfrac{2\pi c_0^3}{15\sqrt{3}} + \hor^8\ell^6 \big(\tfrac{4 c_0^6}{3} - \tfrac{88 c_0^9 \pi^2}{675}\big)\Big)\\
+ {\cal O}(z_{\textrm{cut}}^2) + {\cal O}(T^{12}\ell^{10}) + {\cal O}(\lambda^3/\ell^3)\,.\label{eq:s107}
\end{multline}
The first line recovers the HEE results of \cite{Fischler:2012ca, Erdmenger:2017pfh}. 

The second derivative of the area \eqref{eq:s107} with respect to $\pm\lambda$ evaluated at $\lambda=0$ yields the QNEC quantity $\mathcal{S}''_\pm$ used in the main text.
\begin{multline}
\frac{1}{2\pi}\,\mathcal{S}''_\pm = -\frac{1}{\pi^2c_0^3\ell^4} + \frac{\hor^4\,c_0^3}{15\sqrt{3}\pi} - \hor^8\ell^4\bigg(\frac{44c_0^9}{675}-\frac{2c_0^6}{3\pi^2}\bigg) \\
+{\cal O}(T^{12}\ell^8)
\label{eq:s108}
\end{multline}

This is our main result in the limit of small separations. For comparison with our numerical results in the main text we evaluate \eqref{eq:s108} using \eqref{eq:c0}. (We set $\pi T=1$.)
\eq{
\frac{1}{2\pi}\,\mathcal{S}''_\pm \approx -\frac{0.06498}{\ell^4} + 0.01910 - 0.08289\, \ell^4
}{eq:s109}
The number $0.06498$ reproduces the correct vacuum result \eqref{eq:Sppvac}, while the numbers $0.01910$ and $0.08289$ appear %
in the fit in the inset of the left Fig.~\ref{fig:QNECblackbrane}.

\paragraph{Large width expansion - } If $T\ell\gg 1$ then the holographic depth $z_\ast$ approaches the horizon, 
\eq{
z_\ast = \hor^{-1} (1-\epsilon)\qquad 0<\epsilon\ll 1\,.
}{eq:s14}
This means that we have again a small parameter that we can use for perturbative purposes, namely $\epsilon$. However, a technical difficulty is that integrals like \eqref{eq:s37} now acquire terms that diverge like $\ln\epsilon$ or $1/\epsilon$ due to the behavior of the integrands near the upper integration boundary $x=1$. Thus, we need to isolate these divergences as we expand around $\epsilon=0$. 

We encounter two types of delicate integrals. The first one is of the form
\eq{
I_1[h(x)]=\int\limits_0^1\frac{h(x)\,\extd x}{\sqrt{1-x}\,(1 - x + \epsilon x)^{3/2}} = \frac{2h(1)}{\epsilon} + {\cal O}(\ln\epsilon) %
}{eq:int1}
and the second one reads
\begin{multline}
I_2[h(x)]=\int\limits_0^1\frac{h(x)\,\extd x}{\sqrt{(1-x)(1 - x + \epsilon x)}} = -h(1)\,\ln\frac\epsilon4 \\
+ \int\limits_0^1\extd x \,\frac{h(x)-h(1)}{1-x} - \big(h(1)+h^\prime(1)\big)\,\frac\epsilon2\,\ln\frac\epsilon4 +{\cal O}(\epsilon)
\label{eq:int2}
\end{multline}
where in both cases the function $h(x)$ must be (and in all our cases will be) Taylor-expandable around $x=1$. We have also simple explicit expressions for the subleading terms, but do not display them since we are not going to use them (with one exception). By virtue of the formulas above we now evaluate the three relevant integrals.

Let us start with the integral \eqref{eq:s100}. We rewrite it as
\eq{
\frac{\lambda}{2}=\Lambda\,z_\ast\,I_1[h_\Lambda(x)] 
}{eq:s112}
with $h_\Lambda(1)\simeq 1/(8\sqrt{6})+{\cal O}(\epsilon)$ %
where $\simeq$ denotes equality up to terms of irrelevant order in $\lambda$. Using \eqref{eq:int1} for small $\epsilon$ the integral \eqref{eq:s112} yields
\eq{
\Lambda \simeq  2\sqrt{6}\epsilon\lambda\hor + {\cal O}(\lambda\epsilon^2\ln\epsilon) \,.
}{eq:s110}

The next integral we consider is \eqref{eq:s38}, which determines $\epsilon$ defined in \eqref{eq:s14} in terms of $\ell$ and $\lambda$. Again we slightly rewrite the integral,
\eq{
\frac{\ell+\lambda}{2z_\ast}\simeq I_2[h_z(x)]
+\lambda^2\hor^2 \epsilon\,I_1[h_\lambda(x)]
}{eq:s116}
which for small $\epsilon$ by virtue of \eqref{eq:int1} and \eqref{eq:int2} expands as
\eq{
\frac{\ell+\lambda}{2z_\ast}\simeq-h_z(1)\ln\frac\epsilon4 + h_z^0 + 2\lambda^2 h_\lambda(1) + {\cal O}(\epsilon\ln\epsilon)
}{eq:s119}
with $h_z(1)=2h_\lambda(1)=1/(2\sqrt{6})+{\cal O}(\epsilon)$ and
$h_z^0 %
\approx -0.25032$~\footnote{%
The explicit expression for $h_z^0$ follows from the integral formula \eqref{eq:int2} and reads $h_z^0 = \int_0^1\extd x \,[h_z(x)-h_z(1)]/(1-x)$  with $h_z(x)=x^3/[(1+x)W(x)]$ where $W(x)=\sqrt{(1 + x^2) (1 - x + x^2) (1 + x + x^2)}$.
}, 
yielding
\eq{
\epsilon \simeq \epsilon_0 \exp\big[-\sqrt{6}(\ell+\lambda)\hor+\lambda^2\hor^2\big] + \dots
}{eq:s111}
where the ellipsis refers to terms that are exponentially suppressed as compared to the one displayed. Numerically, $\epsilon_0=4\exp[h_z^0/h_z(1)]\approx1.173487$.

Finally, we evaluate the area integral \eqref{eq:s37}. We split it into $\lambda$-independent and $\lambda$-dependent terms
\begin{multline}
I_{\cal A} \simeq I_2[h_z(x)+ \epsilon\,k_z(x)] \lambda^2\hor^2\epsilon\,I_1[h_\lambda(x) + \epsilon\,k_\lambda(x)] \\ 
\simeq \frac{\ell+\lambda}{2z_\ast} + \epsilon\,\big(I_2[k_z(x)] + \lambda^2\hor^2\epsilon\,I_1[k_\lambda(x)] \big)
\label{eq:s121}
\end{multline}
with the same functions $h_z$ and $h_\lambda$ as in \eqref{eq:s116}, $k_z(1)=-1/2$ and $k_\lambda(1)=-\sqrt{6}/4$. Physically, the reason why the split of the integrals in \eqref{eq:s121} into $h$ and $k$ is useful is related to the fact that for large $T\ell$ HEE scales linearly with $\ell$. 

The integration formulas \eqref{eq:int1} and \eqref{eq:int2} together with the results above yield for the area \eqref{eq:s36}
\begin{multline}
{\cal A} \simeq \frac{1}{z_{\textrm{cut}}^2} + \frac{\ell+\lambda}{z_\ast^3} + \frac{1}{z_\ast^2}\,\big(b_0 + b_1\epsilon + b_{\textrm{\tiny log}}\epsilon\ln\epsilon\big) \\
+ \lambda^2\hor^4\,b_2\epsilon + {\cal O}(z_{\textrm{cut}}^2) + {\cal O}(\epsilon^2\ln\epsilon) 
\label{eq:s42}
\end{multline}
with $b_0\approx -0.66589$, $b_1\approx  -0.08889$~\footnote{%
To evaluate $b_1$ also the first subleading term not displayed in the integral formula \eqref{eq:int2} is needed. The explicit expression for $b_1$ reads
 $b_1 = 1+2k_0(1)-b_{\textrm{\tiny log}}\ln4  + 2\int_0^1\extd x\,[k_1(x)+\frac12]/(1-x) %
 -\int_0^1\extd x\,\int_x^1\extd y\,[k_0(y)-k_0(1)]/(1-y)$
with the functions $k_0(x)=[x^4+x^2+1-W(x)]/[x^3W(x)]$, $k_1(x)=(1-x)(3x^6 + 2 x^5  + 4x^4 + 2x^3 + 4x^2  + 2x + 1)/[2 x^2 (1 + x) (1 + x^2) W(x)]-1/(2x^2)$ and $W(x)=\sqrt{(1 + x^2) (1 - x + x^2) (1 + x + x^2)}$. The explicit expression for $b_0$ follows from the integral formula \eqref{eq:int2},  $b_0 =-1+2\int_0^1\extd x\,k_0(x)$.
}, $b_2=-\sqrt{6}$ and $b_{\textrm{\tiny log}}=\sqrt{6}/2$.
For $\lambda=0$ the area \eqref{eq:s42} establishes a result for HEE,
\begin{multline}
\mathcal{S}_{\textrm{EE}} = \frac{1}{2\pi}\,\Big[\frac{1}{z_{\textrm{cut}}^2} + \ell \hor^3 + \hor^2 \,b_0 
+ e^{-\sqrt{6}\ell\hor} \cdot \\
\cdot \hor^2\,\epsilon_0\, \big(2b_0+b_1+b_{\textrm{\tiny log}}\ln\epsilon_0 \big)\Big]
\label{eq:s23}
\end{multline}
where we neglected terms that vanish as the cutoff is removed, $z_{\textrm{cut}}\to 0$, and terms that are exponentially suppressed like $\ell\exp[-2\sqrt{6}\ell\hor]$. Note that all terms of the form $\ell \exp[-\sqrt{6}\ell\hor]$ cancel. Numerically, the cutoff-independent terms read (setting $\pi T=1$)
\eq{
2\pi\mathcal{S}_{\textrm{fin}} \approx \ell-0.666 - 1.437 \,e^{-\sqrt{6}\ell} + {\cal O}(\ell\,e^{-2\sqrt{6}\ell})\,.
}{eq:s29}
The result above agrees with (5.27) and (B.26) in \cite{Fischler:2012ca}.

The second derivative of the area \eqref{eq:s42} with respect to $\pm\lambda$ evaluated at $\lambda=0$ yields again the QNEC quantity $\mathcal{S}''_\pm$ used in the main text. 
\eq{
\frac{1}{2\pi}\,\mathcal{S}_\pm'' = - \frac{5\sqrt{6}\,\epsilon_0}{4\pi^2}\,\hor^4\,e^{-\sqrt{6}\ell\hor} + \dots
}{eq:s126}
where we neglected terms that are suppressed like $\ell\,\exp[-2\sqrt{6}\ell\hor]$ and used the numerical identity $b_{\textrm{\tiny log}}\ln\epsilon_0 = -2b_0-b_1-b_{\textrm{\tiny log}}$. Note that again all terms of the form $\ell \exp[-\sqrt{6}\ell\hor]$ cancel. Inserting numbers into our large width result \eqref{eq:s126} yields (setting $\pi T=1$)
\eq{
\frac{1}{2\pi}\,\mathcal{S}_\pm'' \approx -0.364053\,e^{-2.44949\ell} \,.
}{eq:s125}
The exponential behavior in \eqref{eq:s125} agrees rather precisely with the numerical data displayed in Fig.~\ref{fig:QNECblackbrane}.

\bibliographystyle{apsrev4-1}
\bibliography{biblio}

\end{document}